# Van Hove Singularity and Lifshitz Transition in Thickness-Controlled Li-Intercalated Graphene

S. Ichinokura[1,*], M. Toyoda[1], M. Hashizume[1], K. Horii[1], S. Kusaka[1], S. Ideta[2,3], K. Tanaka[2], R. Shimizu[4], T. Hitosugi[4], S. Saito[1,5,6], and T. Hirahara[1]

[1]*Department of Physics, Tokyo Institute of Technology, Tokyo 152-8551, Japan*
[2]*UVSOR Facility, Institute for Molecular Science, Okazaki 444-8585, Japan*
[3]*Hiroshima Synchrotron Radiation Center, Hiroshima University, Higashi-Hiroshima 739-8526, Japan*
[4]*Department of Applied Chemistry, Tokyo Institute of Technology, Tokyo 152-8550, Japan*
[5]*Advanced Research Center for Quantum Physics and Nanoscience, Tokyo Institute of Technology, Meguro-ku, Tokyo 152-8551, Japan*
[6]*Materials Research Center for Element Strategy, Tokyo Institute of Technology, 4259 Nagatsuta-cho, Midori-ku, Yokohama, Kanagawa 226-8503, Japan*

(Dated: August 12, 2021)

We demonstrate a new method to control the Fermi level around the van Hove singularity (VHS) in Li-intercalated graphene on the SiC substrate. By angle-resolved photoemission spectroscopy, we observed a clear Lifshitz transition in the vicinity of the VHS by increasing the graphene thickness. This behavior is unexpected in a free-standing Li-intercalated graphene model. The calculation including the substrate suggests that the surface state stabilizes the Fermi level around the VHS of the Dirac bands via hybridization. In addition, we found that a sizable Schottky barrier is formed between graphene and the substrate. These properties allow us to explore the electronic phase diagram around the VHS by controlling the thickness and electric field in the device condition.

Since the demonstration of superconductivity in twisted bilayer graphene [1], the many-body effect when the van Hove singularity (VHS) is tuned at the Fermi level ($E_F$) has been intensively studied in two-dimensional materials [1-11]. In the twisting method, the moiré potential plays a vital role in forming a flat band. Another route to tune the VHS at $E_F$ in graphene systems is heavy carrier doping by intercalation of guest metals [12-16]. It pulls down the native flat band around the saddle point (SP) of π-band, initially located ~2 eV above $E_F$, into an occupied state. Variety of electronic phase such as spin-density-wave and unconventional superconductivity is predicted around VHS due to the enhancement of the strong correlation effect [17-21]. So far, in Ca- [12], Gd- [16], Cs- [13], or Yb- [14,15] intercalated graphene, the sign of the many-body effect is seen as a renormalization of the band structure. It is called an extended VHS [12-16], where the flat band is pinned to $E_F$ in a wide range of wavenumber space [12,14,15]. However, $E_F$ tuning in the vicinity of the VHS has been achieved only by adsorbing another element [12,14] or heating treatment [15]. Compared to the twisted systems processed in field-effect transistor devices, the poor controllability of the intercalation system caused a lack of macroscopic evidence of the enhanced correlation effect with transport measurements.

It is known that Li-intercalated bilayer graphene can be fabricated from epitaxial monolayer graphene on SiC by Li deposition [22-25]. Here, the buffer layer is lifted from the substrate and becomes the bottom graphene layer [26]. The resulting structure is $C_6LiC_6$/Li-terminated SiC, as shown in Figs. 1(a) and (b). The √3×√3-R30° superlattice of Li causes a periodic modulation called Kekulé-O-type, which breaks the chiral symmetry of graphene, resulting in a gapped double Dirac cone [23,27]. However, the detailed electronic structure around $E_F$ has not yet been clarified. Especially, there are no reports on the VHS and related band-renormalization. Also, the band structure of multilayer Kekulé-ordered graphene has not been elucidated.

In this Letter, we investigated the thickness dependence of the band structure of Li-intercalated graphene (LIG) with angle-resolved photoemission spectroscopy (ARPES) and density functional theory (DFT) calculations. In the bilayer LIG/SiC, we observed a flat band at $E_F$, indicating the extended VHS. The DFT calculations for free-standing LIG expects that the SP in multilayer can be far below $E_F$ due to the interlayer interaction. However, we found that the SP robustly stays near $E_F$ in the actual multilayer systems. As the thickness is increased, the doping level is systematically controlled, and a Lifshitz transition takes place. We performed DFT calculations for LIG/SiC including the substrate. The surface electronic state of Li-terminated SiC hybridizes with the Dirac band dispersion near the Fermi surface, and this possibly contribute to the stabilization of VHS near $E_F$. It was also found that a sizable Schottky barrier is formed between LIG and the substrate. This indicates the capability of the SiC substrate as a back gate, which is promising to explore the transport properties under the correlation effect by the field-effect tuning of $E_F$ in thickness-controlled LIG.

We prepared epitaxial graphene on the surface of an n-type Si-rich 4H-SiC(0001) single crystal by thermal decomposition in an Ar atmosphere. The thickness of graphene is controlled by optimizing the heating condition. The resulting thickness is evaluated from band dispersion at the $\bar{K}$ point, as shown in supplementary material 1(SM-1). Li was deposited on graphene at room temperature from a resistively heated dispenser in an ultrahigh vacuum chamber. The intercalation processing can be monitored in the electron-diffraction pattern. As written in SM-2, an $n-1$ layer graphene/buffer-terminated SiC turns into an $n$ layer LIG/Li-terminated SiC. Here, we use $n$ ($n$ = 2, 3, 4, and 5) to describe the number of layers. We most likely think that the approximate structure of $n$ layer LIG is $(C_6Li)_{n-1}C_6$. We continued the deposition of Li until the intensity of $\sqrt{3}\times\sqrt{3}$-Li spots saturated to minimize the Li vacancy. ARPES measurements were performed *in situ* after the sample preparation with a commercial hemispherical photoelectron spectrometer equipped with angle and energy multidetections. We used two different apparatuses: Scienta Omicron R4000 in the lab with unpolarized HeIα (21.2 eV) radiation and MBS A1 at BL-7U of UVSOR-III using *p*- or *s*-polarized photons in the energy range of 14-40 eV [28]. The measurements were performed at room temperature in the lab and at 13 K in UVSOR (the data shown are taken at 13 K unless otherwise indicated).

In the $\sqrt{3}\times\sqrt{3}$-Brillouin Zone (BZ), the $\bar{\Gamma}$ and $\bar{K}$ points of the 1×1-BZ are equivalent. Thus, we focus on the $\sqrt{3}\times\sqrt{3}$-BZ in this paper. The band structure of $C_6LiC_6$/Li-SiC along the $\bar{\Gamma}$ - $\bar{M}_{\sqrt{3}}$ line is shown in Fig. 1(c). $\pi_1$, $\pi_1^*$, $\pi_2^*$, SS, and SiC bands are seen. SS and SiC bands are derived from the surface and bulk state of the substrate, respectively, as described later. A pair of Dirac cone $\pi_1^*$ and $\pi_1$ has a gap of 0.37 eV, the same as reported in Ref. [23]. It is noteworthy that the $\pi_2^*$, the other Dirac band, has a flat dispersion near the Fermi energy at the $\bar{M}_{\sqrt{3}}$ point. As shown in the series of electron distribution curves (EDC) in Fig. 1(d), the flat band spans a wide range of wavenumbers; from 0.66 to 0.97 Å$^{-1}$, the bandwidth is less than ten meV. Such a flatness over a wide wavenumber range is evidence of the extended VHS [12-16]. Since the flat band originates from the SP of graphene at the $\bar{M}$ point, it is electron-like in the orthogonal wavenumber axis. Fig. 1(e) shows that the bottom of the parabolic $\pi_2^*$ band is located on $E_F$.

The Fermi surface is depicted in Fig. 1(f). In the first $\sqrt{3}\times\sqrt{3}$ BZ, both $\pi_1^*$ and $\pi_2^*$ approximately have a small and large hexagram contour centered at the $\bar{\Gamma}$ point (see the blue and yellow guidelines, respectively). Considering the BZ folding, the $\pi_1^*$ hexagram is originated from triangular electron pockets at the $\bar{K}$ and $\bar{K}'$ point of the 1×1 BZ. $\pi_2^*$ is a large circular hole pocket centered at the $\bar{\Gamma}$ point. From the volume of the Fermi surface, the carrier densities of $\pi_1^*$ and $\pi_2^*$ bands are estimated to be $1.4 \times 10^{14}$ cm$^{-2}$ and $3.5 \times 10^{14}$ cm$^{-2}$, respectively. The total is $4.9 \times 10^{14}$ cm$^{-2}$, roughly consistent

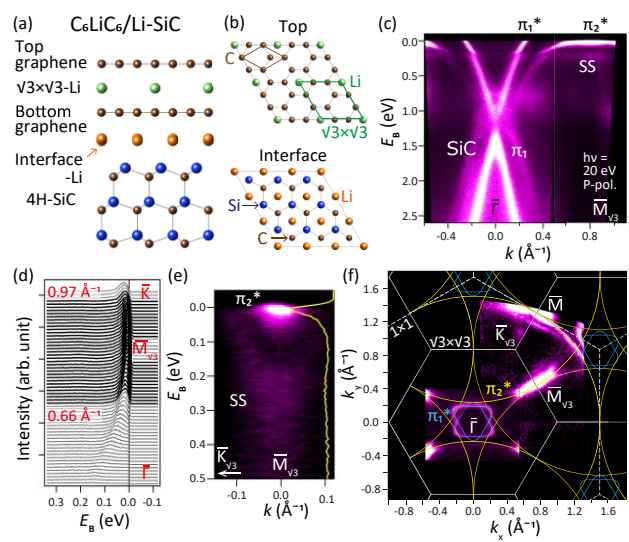

**FIG. 1** (a)(b) Top (a) and side (b) view of the atomic model of $C_6LiC_6$/Li-SiC. (c)-(f) High-resolution ARPES data taken with *p*-polarized light at $hv$ = 20 eV. (c) Band dispersion and (d) EDC mapping along the $\bar{\Gamma}$ - $\bar{M}_{\sqrt{3}}$ direction. $\pi_1$, $\pi_1^*$, $\pi_2^*$, SS and SiC bands are seen. SS and SiC bands are derived from surface and bulk state of the substrate, respectively. (e) Band dispersion along $\bar{M}_{\sqrt{3}}$ - $\bar{K}_{\sqrt{3}}$ direction. The yellow line is the EDC at $k$ = 0. (f) Fermi surface. The white dashed and solid lines indicate the boundary of 1×1 and $\sqrt{3}\times\sqrt{3}$ BZ, respectively. The blue and yellow solid lines are the guideline for the contour of $\pi_1^*$ and $\pi_2^*$ band.

with the atomic density of $\sqrt{3}\times\sqrt{3}$-Li ($6.3\times 10^{14}$ cm$^{-2}$). This implies that most of the electrons of $\sqrt{3}\times\sqrt{3}$-Li between graphene are transferred to the two graphene layers. There are also features near the $\bar{M}_{\sqrt{3}}$ point as discussed later.

To see what changes when we start from multilayer graphene, we first investigated the evolution of the band structure of $(C_6Li)_{n-1}C_6$ upon the thickness $n$ using DFT calculations. In Figs. 2(a)-(d), we show the calculated band structure of $(C_6Li)_{n-1}C_6$ for $n$ = 2, 3, 4, and 5. In all the models, the stacking manner of graphene is AA. Li occupies the same hollow site between the graphene layers, i.e., Li atoms are aligned vertically [29,30]. The computational detail is described in SM-3. The same number of Dirac bands as graphene layers are seen due to the interband interaction. These multiple bands can be roughly divided into two groups: electron-like bands around the $\bar{\Gamma}$ point and hole-like bands around the $\bar{M}_{\sqrt{3}}$ point. For $n$ = 2, the electron and hole bands agree with $\pi_1$, $\pi_1^*$ and $\pi_2^*$ in Fig. 1(c). Even for $n \geq 3$, all the bands have a gap at the Dirac point, indicating that the chiral symmetry breaking is also effective in the multilayer case. The carbon layers are labeled as shown in the inset of each figure, and in Figs. 2 (b)-(d), the band structure is classified

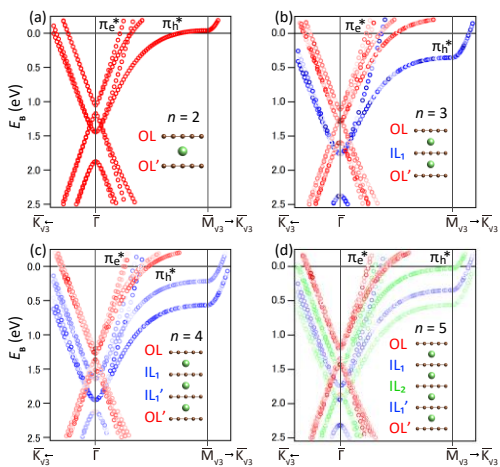

**FIG. 2** (a)-(d) Band structure of $(C_6Li)_{n-1}C_6$ with $n = 2$, 3, 4, and 5 obtained by DFT calculations. The inset shows the schematic structure. The outer, the first inner, and the second inner carbon layers are labeled as OL, OL', $IL_1$, $IL_1'$, and $IL_2$. In (b)-(d), the band structure is classified by the wave function distribution; states that are localized at OL and OL' are shown in red, those at $IL_1$, $IL_1'$ are shown in blue, and those at $IL_2$ are shown in green.

by the wave function distribution; states that are localized at the outer layers OL and OL' are shown in red, those at the first inner layers $IL_1$, $IL_1'$ are shown in blue, and those at the second inner layer $IL_2$ are shown in green. We find that the electron bands originate mainly from the surface graphene (OL, OL'), while the hole bands originate from the inner graphene ($IL_1$, $IL_1'$, $IL_2$). Here, OL and OL' ($IL_1$ and $IL_1'$) are equivalent because of the spatial inversion symmetry in the surface-normal direction. For the comparison with the experiments, we will refer to the electron and hole pockets closest to $E_F$ as $\pi_e^*$ and $\pi_h^*$, respectively. The VHS ($\pi_h^*$) is generally not pinned at $E_F$ in the multilayer case because the interlayer interaction mainly determines the band structure. For example, for $n = 3$, the SP of $\pi_h^*$ is 0.4 eV below $E_F$ at the $\overline{M}_{\sqrt{3}}$ point.

Figures 3(a)-(l) show the experimental band structures and the Fermi surfaces of $n$ layer LIG $(C_6Li)_{n-1}C_6$/Li-SiC with different thicknesses. Here, $2 < n < 3$ and $3 < n < 4$ means a mixture of bilayer/trilayer and trilayer/quadlayer, respectively (See SM-1, 2). The bands look pretty similar to that of $C_6LiC_6$/Li-SiC shown in Fig. 1. The most clearly resolved features are the electron band $\pi_1^*$ and the hole band $\pi_2^*$. The calculations in Fig. 2 showed that for $n \geq 3$, there are two electron bands originated from OL and OL'. It is natural to assign the observed $\pi_1^*$ to the one from the topmost outer layer because of the surface sensitivity of ARPES. In agreement with the calculation, $\pi_1$ is a massive Dirac cone(Figs. 3(a), (c), (e)). By the numerical fittings described in SM-4, we found that both the Dirac point (DP) and top of the $\pi_1$ band shift to the higher binding energy. The difference between them was

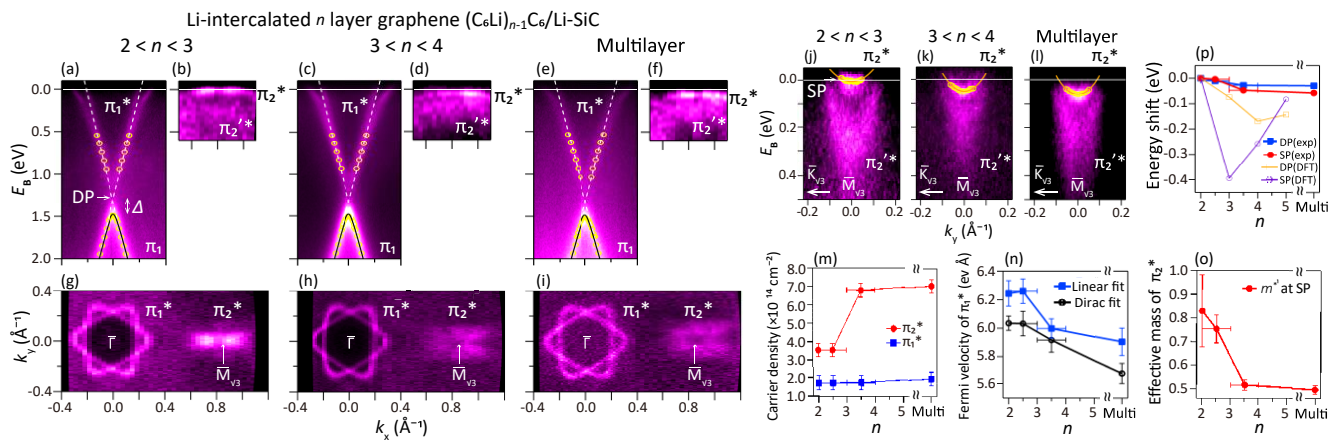

**FIG. 3** Evolution of the band structure of Li-intercalated $n$ layer graphene $(C_6Li)_{n-1}C_6$/Li-SiC. (a)-(f) Band dispersion along the $\overline{\Gamma}$-$\overline{M}_{\sqrt{3}}$ direction. (a)(c)(e) and (b)(d)(f) show the data at the vicinity of the $\overline{\Gamma}$ and $\overline{M}_{\sqrt{3}}$ points, respectively. The yellow opened circles represent the peak positions obtained by Lorentzian fitting to the raw ARPES spectra. The dashed white line is a linear fit of the dispersion. The solid black line is a fit by the massive Dirac cone equation to $\pi_1$(See SM-4). (g)-(i) Fermi contour. (j)-(l) Band dispersion along the $\overline{M}_{\sqrt{3}}$-$\overline{K}_{\sqrt{3}}$ direction. The yellow open circles are the same as (a)(c)(e). The solid orange line indicates the parabolic fit to the peak positions. (m) Thickness dependence of the carrier density calculated from the volume of the Fermi surface. (n) Thickness dependence of the Fermi velocity of $\pi_1^*$. (o) Thickness dependence of the effective mass of $\pi_2^*$. (p) Thickness dependence of the energy shift of the DP and SP. DP of $\pi_1^*$ and SP of $\pi_2^*$ are evaluated from the linear and parabolic fitting, respectively. DP of $\pi_e^*$ and SP of $\pi_h^*$ is directly obtained from the DFT data. All the data were taken with $h\nu = 21.2$ eV at room temperature.

0.16 ± 0.01 eV irrespective of $n$, meaning no significant change in the gap size. On the other hand, the Fermi velocity of $\pi_1^*$ slightly decreased with increasing $n$, as shown in Fig. 3(n).

The hole bands including $\pi_h^*$ in DFT were observed as $\pi_2^*$ and $\pi_2'^*$ bands in the experiments (Figs. 3(b), (d), (f)). Same as $\pi_h^*$, the SP of $\pi_2^*$ is located below the Fermi level for $n > 3$. Accordingly, as shown in Figs. 3(g)-(l), the system exhibits a Lifshitz transition upon increasing thickness. The $\pi_2^*$ Fermi surfaces touch each other at the $\overline{M}_{\sqrt{3}}$ points for $n < 3$, while they become separated for $n > 3$. We evaluated the carrier density of $\pi_1^*$ and $\pi_2^*$, as shown in Fig. 3(m). $\pi_1^*$ slightly expands due to the slight shift of the DP and decrease of the Fermi velocity. In contrast, $\pi_2^*$ shows a discontinuous expansion with the formation of the third layer and the Lifshitz transition.

We performed a parabolic fit on $\pi_2^*$ along the $\overline{M}_{\sqrt{3}}$- $\overline{K}_{\sqrt{3}}$ direction, as shown in Figs. 3(j)-(l). Figure 3(o) depicts the effective mass of $\pi_2^*$, which decreases when the VHS moves below $E_F$. We plotted the energy shift of the DP(SP) of experimental $\pi_1^*(\pi_2^*)$ and theoretical $\pi_e^*(\pi_h^*)$ as a function of the thickness (Fig. 3(p)). One can see that, in calculation results, both DP and SP change by sub-eV. On the other hand, the experimental shifts of DP and SP are only 30 and 60 meV, respectively. This suggests that the large shift of SP by the interlayer interaction is suppressed in the actual LIG. The thickness-dependent Lifshitz transition was caused by the small shift of SP, resulting from an interplay between the interlayer interaction and the suppression effect, as discussed below.

Intuitively, the condition that the VHS is pinned at $E_F$ is robust against slight electron doping due to the large density of state. Many experimental studies demonstrated the pinning of the flat band to $E_F$ is assisted by the many-body effect[12,14]. However, within the single-particle picture, the hybridization with the substrate's electronic states also significantly impacts the flat band and VHS [13]. Figure 1(c) shows some pieces of evidence of the substrate effect: the SS band, SiC band, and kink structures in $\pi_1^*$ and $\pi_2^*$. They are more clearly seen by changing the energy and polarization of the incident light (See SM-5). Figure 4(a) shows the superposition of the spectra measured on $C_6LiC_6$/Li-SiC with $s$- and $p$-polarized light at $hv = 18$ eV. The electron-like band (solid arrows) and the hole-like bands (dotted arrows) are observed. These structures are more accentuated by the second derivative, as shown in Fig. 4(b), indicating that the kink and the electron-like band are originated from the hybridization of two bands. We performed DFT calculations based on the structural model incorporating the substrate and interfaces as shown in Figs. 1(a) and (b). To reduce calculation cost, we approximated the unit cell into $2\sqrt{3}\times2\sqrt{3}$-R30° of graphene. Detail of the analysis is described in SM-3.

The calculated results are shown in Figs. 4(c) and (d). The $\pi_1^*$ and $\pi_2^*$ bands are mainly composed of top and bottom

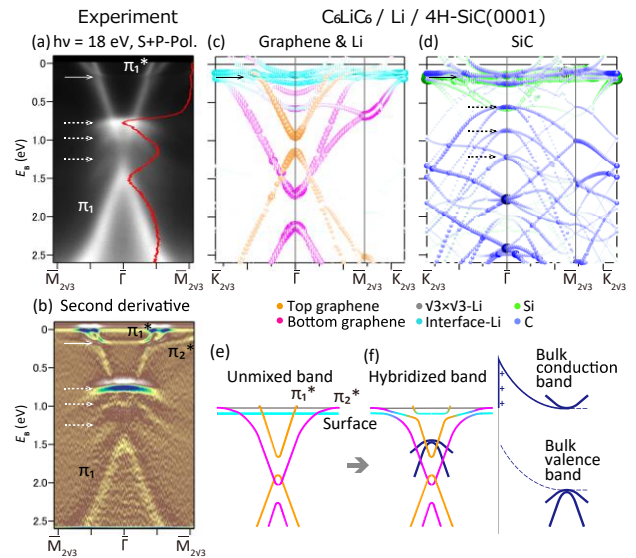

**FIG. 4** (a)Sum of the ARPES intensity taken with $p$- and $s$-polarized light at $hv = 18$ eV around the $\overline{\Gamma}$ point. The solid red line is the EDC at the $\overline{\Gamma}$ point. (b)Second derivative of (a). (c) and (d)The band structures calculated by DFT based on the model shown in Fig 1. (a) and (b). The band structure is projected on C in top/bottom graphene(yellow/magenta), Li in $\sqrt{3}\times\sqrt{3}$/interface(grey/cyan), and Si/C(green/blue) in SiC. Both size and color of the dots are proportional to the projection of those atomic characters. In (a)-(d), Solid and dashed arrows point electron- and hole-like bands, corresponding to the surface and bulk band, respectively. (e) and (f) Schematic illustration of the band hybridization and bending.

graphene, respectively. One characteristic feature is the localized band at 0.2 eV, pointed by solid black arrows. The main component is the surface state of Li-terminated SiC. Here, hybridization occurs at the crossing point of the Dirac cones and the localized bands. Figures 4(e) and (f) schematically illustrate the hybridization of those bands. It qualitatively explains the shape of the electron-like band in the experimental data. The localized nature of the surface state at 0.2 eV is consistent with the SS band seen in Fig. 1(c). This observation suggests that the SiC substrate and the interface Li significantly modify the band structure of $(C_6Li)_{n-1}C_6$ near $E_F$. We suspect that this hybridization effect suppresses the shift of SP even in $n > 2$ from sub-eV to several tens of meV [31].

The hole-like features around 0.7 eV are reproduced in Fig. 4(d) (pointed by black dashed arrows). Judged from the shape, they are the bulk bands of SiC (See SM-6). Here, the valence band top should be located at around 3.2 eV deep inside the bulk since the substrate used in this study is degenerate. This means that considerable band bending (~2.5 eV) occurs from the bulk towards the surface, forming a very high Schottky barrier. The magnitude of the bending is consistent with the reported core level shift, interpreted to originate from the dipole layer at the Li-Si bonding [32]. Therefore, SiC substrate can be used as a back gate to tune $E_F$ of LIG by applying gate voltage. In addition, it should be noted that LIG can be produced by electrochemical intercalation [33,34]. We propose that Li/electrolyte/graphene/buffer-SiC is a promising device, where electrical control of both Li-intercalation and tuning of $E_F$ is possible as reported for other systems [35].

In conclusion, we succeed in the $E_F$ tuning in $(C_6Li)_{n-1}C_6$/Li-SiC by the thickness control, evidenced by the Lifshitz transition. The hybridization of the Dirac cones with the surface state of Li-terminated SiC seems to make it possible to control $E_F$ in the vicinity of the VHS. We also found that Li-intercalation between LIG and SiC naturally forms a sizable Schottky barrier. These properties suggest that LIG on SiC is an ideal platform of the field-effect transistor to solve the long-standing problem related to the VHS.

This work has been supported by Grants-In-Aid from Japan Society for the Promotion of Science (No 18H03877, 20H05183, 19K15443, 21K14533, 19H01823), the Murata Science Foundation (No. H30-084, H30-004), the Asahi Glass Foundation, the Sumitomo Foundation, Tokyo Tech. Challenging Research Award, and Tokyo Tech Advanced Researchers (STAR). The ARPES measurements were performed under the UVSOR proposal No. 19-858, 20-777. A part of this work was conducted at NanofabPF, Tokyo Tech, supported by "Nanotechnology Platform Program" of the Ministry of Education, Culture, Sports, Science and Technology (MEXT), Japan, Grant Number JPMXP09F20IT0008. This work was partly supported by MEXT Elements Strategy Initiative to Form Core Research Center through Tokodai Institute for Element Strategy, Grant Number JPMXP0112101001.

*Electronic address: ichinokura@phys.titech.ac.jp

Supplementary Material:

# Van Hove Singularity and Lifshitz Transition in Thickness-Controlled Li-Intercalated Graphene

S. Ichinokura[1,*], M. Toyoda[1], M. Hashizume[1], K. Horii[1], S. Kusaka[1], S. Ideta[2,3], K. Tanaka[2], R. Shimizu[4], T. Hitosugi[4], S. Saito[1,5,6], and T. Hirahara[1]

[1]*Department of Physics, Tokyo Institute of Technology, Tokyo 152-8551, Japan*
[2]*UVSOR Facility, Institute for Molecular Science, Okazaki 444-8585, Japan*
[3]*Hiroshima Synchrotron Radiation Center, Hiroshima University, Higashi-Hiroshima 739-8526, Japan*
[4]*Department of Applied Chemistry, Tokyo Institute of Technology, Tokyo 152-8550, Japan*
[5]*Advanced Research Center for Quantum Physics and Nanoscience, Tokyo Institute of Technology, Meguro-ku, Tokyo 152-8551, Japan*
[6]*Materials Research Center for Element Strategy, Tokyo Institute of Technology, 4259 Nagatsuta-cho, Midori-ku, Yokohama, Kanagawa 226-8503, Japan*

[*] Electronic address: ichinokura@phys.titech.ac.jp

**Contents**

SM-1. Growth of graphene with thickness control
SM-2. Li-intercalation
SM-3. Computational details
SM-4. Analysis of thickness dependence of Li-intercalated graphene
SM-5. Incident light dependence
SM-6. Bulk band of SiC

# SM-1. Growth of graphene with thickness control

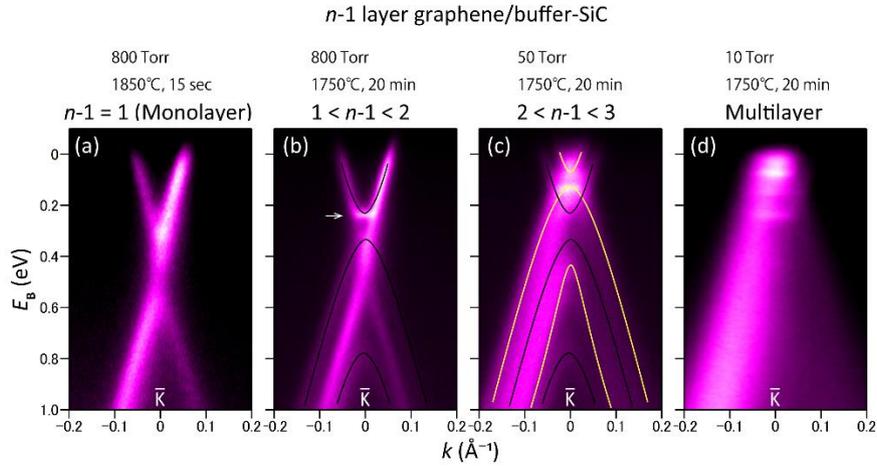

**Figure S1** (a)-(d) ARPES intensity of the as-grown graphene with different thicknesses around $\overline{\mathrm{K}}$ point. The growth condition (pressure of Ar gas, heating temperature and time) is shown above the figures. The solid black curves in (b) and (c) are the guideline for the dispersion of the bilayer graphene. The solid yellow curves in (c) are the guideline for the dispersion of the trilayer graphene. All the data were taken at $hv = 21.2$ eV at room temperature.

We prepared epitaxial graphene on the surface of an n-type Si-rich 4H-SiC(0001) single crystal by thermal decomposition with direct Joule heating in Ar atmosphere. The thickness of graphene is controlled by optimizing the heating temperature(1750-1850°C), time(15sec-20min), and Ar pressure(10-800 Torr). Figure S1 shows the band dispersion of graphene with the various layers used in this study. Here, we count the thickness as '$n$-1 layer' ($n$ =2, 3, 4, and 5) since it becomes $n$ layer Li-intercalated graphene after Li deposition, as described SM-2. (a) is identified as monolayer graphene because it has a typical single Dirac cone. In (b), the second layer is partially formed. The solid black line emphasizes the dispersion of bilayer graphene. Slightly below 0.2 eV, a flat band characteristic for the mixture of monolayer and bilayer graphene is seen as pointed by the white arrow[1]. Same as (b), (c) is the mixture of bilayer and trilayer graphene[2]. The yellow solid line emphasizes the dispersion of the trilayer graphene. In (d), the band structure is that of multilayer graphene: a broad Dirac band whose Dirac point locates near the Fermi level[2].

**SM-2. Li-intercalation**

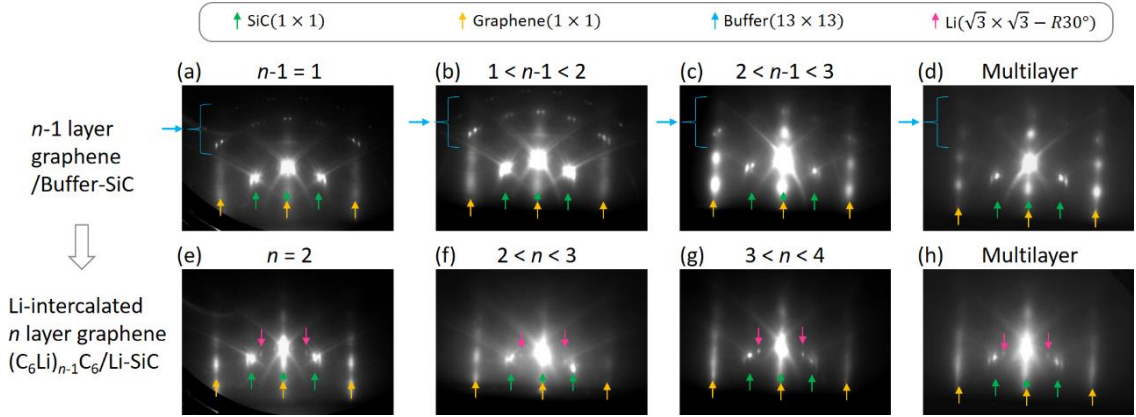

**Figure S2** RHEED pattern of epitaxial graphene on SiC(0001) (a)-(d)before and (e)-(h)after the deposition of Li, with various thicknesses. The diffraction spots of SiC(1×1), graphene(1×1), buffer(13×13), and Li($\sqrt{3}\times\sqrt{3}$) are pointed by green, yellow, blue, and pink arrows, respectively.

The intercalation processing can be observed as the change of 2D periodic structure seen in the reflection-high-energy-electron-diffraction (RHEED) pattern. Figure S2(a) shows the RHEED pattern of the pristine monolayer graphene. The fundamental spots (1×1) of both the graphene and the surface of the SiC substrate are seen. The superlattice spots (13×13 for graphene, $6\sqrt{3}\times6\sqrt{3}$-R30° for SiC(0001)) correspond to the lattice mismatch between graphene and the SiC substrate. Here, we have a buffer layer between the graphene and the SiC. 30% of the carbon atoms in the buffer layer form the $C_{buffer}$-Si covalent bond[3,4]. The $C_{buffer}$-Si bond has the lateral distribution the same as the lattice mismatch between buffer and SiC, makes the strong periodic potential, which can be seen as very bright spots of 13×13. As Li deposition, the intensity of the 13×13 diffraction spots is dramatically weakened, in agreement with the previously reported change of the LEED pattern[5]. According to the STM[5,6] and ARPES[7] studies, the fainting of diffraction spots is caused by breaking the $C_{buffer}$-Si bonds due to the Li-intercalation between the buffer layer and SiC. It saturates because Li terminates the SiC, which turns the buffer layer into graphene. Thus, epitaxial monolayer graphene turns into quasi-free standing bilayer graphene(QFBLG). Further deposition of Li leads to the formation of $\sqrt{3}\times\sqrt{3}$-R30°, as shown in Figure S2(e). As reported by the LEED study previously[5], this indicates that QFBLG turns into the Li-intercalated bilayer graphene, $C_6LiC_6$. Figures S2(b)-(d) show the RHEED pattern of the thicker pristine graphenes. The same periodic structures as the monolayer case are seen. As the graphene has more thickness, graphene 1×1 gets more three-dimensional character, and 13×13 spots get weakened. However, after Li deposition, the same change of the RHEED pattern as monolayer case (from Fig. S2(a) to (e))was seen for all the thickness. The disappearance of 13×13 spots and growth of $\sqrt{3}\times\sqrt{3}$-R30° spots are seen in Figs. S2(f)-(h). This indicates that "*n*-1 layer graphene/buffer-terminated SiC" turns into "$n$ layer Li-intercalated graphene/Li-terminated SiC". Since √3×√3-R30° periodicity was reproducibly seen, we estimate the approximate structure of $n$ layer Li-intercalated graphene is $(C_6Li)_{n-1}C_6$.

**SM-3. Computational details**

The electronic structure calculations are performed by using QUANTUM ESPRESSO package [8,9] within the framework of the Kohn-Sham density functional theory (DFT) [10,11]. The calculations were performed in three steps: (1) model construction, (2) structural optimization, and (3) band structure calculations. In the model construction and the structural optimization steps, the ultra-soft pseudopotentials [12] are used and the valance-electron wavefunctions and charge densities are expanded in a plane-wave basis set with cutoff energies of 40 Ry (wavefunctions) and 400 Ry (charge). In the band structure calculation step, the norm-conserving pseudopotentials [13] are used and the cutoff energies for valence states are 120 Ry (wavefunction) and 480 Ry (charge). The electron-electron interaction is described within the local density approximation (LDA) with the Perdew-Zunger exchange-correlation energy functional [14]. An 8x8x1 k-point sampling mesh is used for integration over the Brillouin zone. In the model construction step, we performed the total energy comparison between different stacking patterns of the graphene layers, the Li atoms, and the substrate. We confirmed that the most stable structures have the AA-stacking graphene layers with the Li atoms intercalated at the hollow site of the graphene as shown in the insets of Figs. 2(a)-(d). For example, when $n$=2, AA-stacking is more favorable than AB-stacking by 0.17 eV/f.u.

To construct the model structure of the bilayer Li-intercalated graphene (LIG) on SiC substrate, $C_6LiC_6$/Li/4H-SiC(0001), we used the lattice matching of 2√3×2√3 graphene and 3×3 SiC(0001). Although it is actually 13×13 graphene and 6√3×6√3 SiC(0001) as seen in the diffraction pattern, we used the smaller cell for the sake of reducing the computational cost. We first performed the structural optimization of the bulk 4H-SiC, and then constructed the 4H-SiC(0001) slab structure. The $C_6LiC_6$ comes to the Si-surface of the substrate while the C-surface is terminated by hydrogen atoms. We performed the total energy comparison to determine the positions of the (1×1) Li adatoms on the Si-surface of 4H-SiC(0001). As shown in Tab. S1, it turns out that the most favorable site is the hollow site of SiC hexagons. We then performed the total energy comparison again to determine the position of $C_6LiC_6$ on the Li adatoms (interface Li). Then, we performed the structural optimization of these models. The details of the optimized structure of $C_6LiC_6$/Li/4H-SiC(0001) are shown in Figs. 1(a) and (b). As for the free-standing LIG $(C_6Li)_{n-1}C_6$, the 2d lattice parameters and the atomic positions are relaxed until the maximum force acting on atoms becomes less than $10^{-6}$ a.u. while the dimension along c-axis is kept constant at 40 Å

(corresponds to the vacuum spacing longer than 20 Å). In the calculations of $C_6LiC_6$/Li/4H-SiC(0001), the atomic positions are relaxed with the same criteria while the 2d lattice parameters are fixed at the same values as those of bulk 4H-SiC and the dimension along c-axis is kept constant at 40 Å (the vacuum spacing is longer than 20 Å). Finally, we performed the band dispersions of these systems and analyzed them into the atom-decomposed band structure by taking the projection at each atomic site.

Table S1: Relative total energy of Li/4H-SiC(0001) with different Li positions.

| Li site | hollow | ontop of Si | ontop of C |
|---|---|---|---|
| ΔE (eV/Li) | 0 | 0.64 | 14.46 |

**SM-4. Analysis of thickness dependence of Li-intercalated graphene**

Figures 3(a)(c)(e) and (j)(k)(l) in the main text are a series of the dispersion of $\pi_1/\pi_1^*$ band at $\bar{\Gamma}$ point and of $\pi_2^*$ band at $\bar{M}_{\sqrt{3}}$ points, respectively. From $\pi_1$ and $\pi_1^*$, we evaluated the energy of the Dirac point (DP), the gap of the Dirac cone (Δ), and the Fermi velocity $v_F$. From $\pi_2^*$, we evaluated the saddle point (SP) energy and effective mass $m_e$. The analysis of the values is conducted by the following method. The DP is defined by the crossing point of the linear functions fitted to the linear part of $\pi_1$ and $\pi_1^*$ Dirac cone. The gap Δ is defined as the energy difference between DP and the top of the lower Dirac cone $\pi_1$. Δ can be estimated by the fitting of the function,

$$E = E_{DP} - \sqrt{\Delta^2 + (v_F k)^2} \qquad (\text{Eq.1}),$$

to the $\pi_1$ band, where the binding energy of Dirac cone $E_{DP}$ is fixed to the value obtained from the linear fitting. Here, $v_F$ is the Fermi velocity, which corresponds to the gradient of the Dirac cone outside of the gap. $v_F$ estimated from two fittings (linear and using Eq.1) are plotted in the same axis in Fig. 3(n). The saddle point (SP) is defined by the bottom of the $\pi_2^*$ band at $\bar{M}_{\sqrt{3}}$ point. Effective mass of $\pi_2^*$ band is estimated by fitting to the parabolic function

$$E = E_{SP} + \frac{(\hbar k)^2}{2m_e m^*} \qquad (\text{Eq. 2}),$$

where $E_{SP}$ is the binding energy of the saddle point, $\hbar$ is the Dirac constant, and $m_e$ is the electron mass.

## SM-5. Incident light dependence

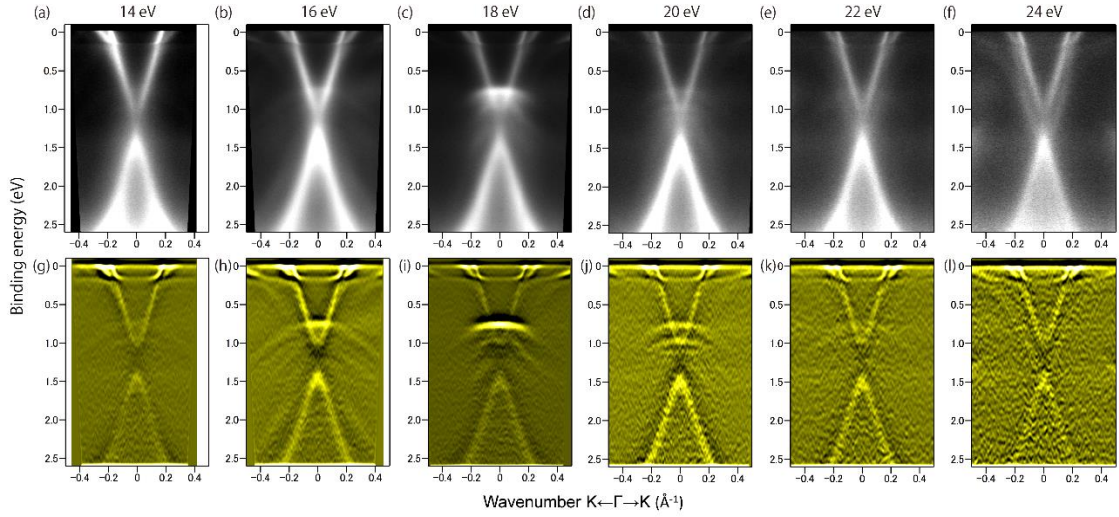

**Figure S3** (a)-(f) Sum of the ARPES intensity of $C_6LiC_6$/Li-SiC around $\bar{\Gamma}$ point taken by *p*- and *s*-polarized light at $h\nu$ = 14-24 eV at 13 K. (g)-(i) Second derivative of the upper panels.

We performed photon-energy-dependent ARPES measurements for *p*- and *s*-polarized light. In the upper and lower line of Fig. S3, we paneled the ARPES intensity maps and their second derivative from 14 to 24 eV, respectively. Here, intensity for *p*- and *s*-polarized light is summed up. One can find electron- and hole-like features at 0.2 and 0.75 eV at the $\bar{\Gamma}$ point, respectively. As seen in the second derivative, the electron-like band anti-crosses with the Dirac bands.

**SM-6. Bulk band of SiC**

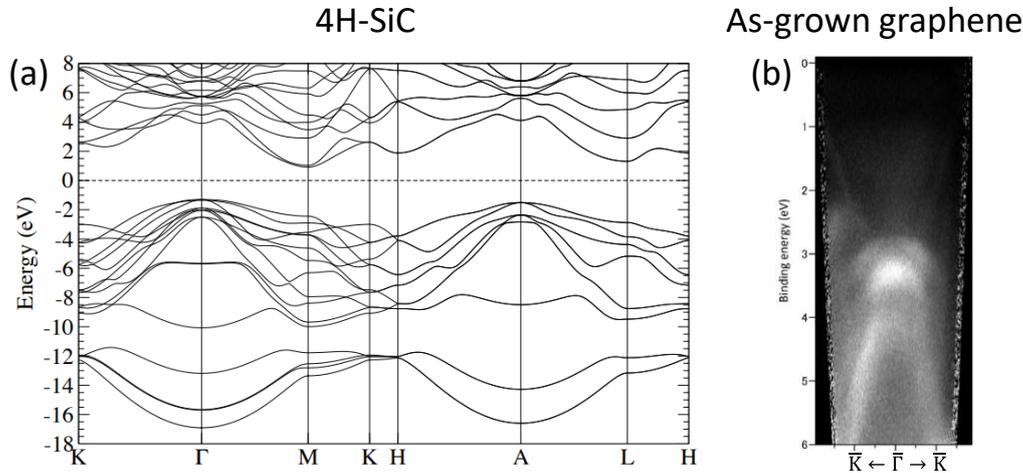

Figure S4 (a)Band structure of 4H-SiC calculated by DFT. (b) Band structure of as-grown monolayer graphene/buffer-SiC around $\bar{\Gamma}$ point measured by ARPES. The data was taken at $h\nu = 21.2$ eV at room temperature

Figure S4(a) shows the band structure of 4H-SiC calculated by DFT. The top of the valence band locates at $\bar{\Gamma}$ point. Figure S4(b) shows the band structure of as-grown monolayer graphene measured by ARPES. The observed hole-like band at $\bar{\Gamma}$ point could be interpreted as the bulk band of SiC, compared with Fig. S4 (a). The top of the valence band is at 2.7 eV.